\documentclass[aps,preprint,showpacs, showkeys]{revtex4}%
\usepackage{amsfonts}
\usepackage{amsmath}
\usepackage{amssymb}
\usepackage{graphicx}%
\providecommand{\U}[1]{\protect\rule{.1in}{.1in}}

\begin{document}
\title[ ]{Possible treatment of the Ghost states in the Lee-Wick Standard Model}
\author{Abouzeid M. Shalaby}
\email{amshalab@ mans.edu.eg}
\affiliation{Physics Department, Faculty of Science, Mansoura University, Egypt.}
\keywords{non-Hermitian models, $\mathcal{PT}$-Symmetric theories, gauge Hierarchy,
Lee-Wick standard model.}
\pacs{12.90.+b, 12.60.Cn, 11.30.Er}

\begin{abstract}
Very recently, the Lee-Wick standard model has been introduced as a non-SUSY
extension of the Standard model which solves the Hierarchy problem. In this
model, each field kinetic term attains a higher derivative term. Like any
Lee-Wick theory, this model suffers from existence of Ghost states. In this
work, we consider a prototype scalar field theory with its kinetic term has a
higher derivative term, which mimics the scalar sector in the Lee-Wick
Standard model. We introduced an imaginary auxiliary field to have an
equivalent non-Hermitian two-field scalar field theory. We were able to
calculate the positive definite metric operator $\eta$ in quantum mechanical
and quantum field versions of the theory in a closed form. While the
Hamiltonian is non-Hermitian in a Hilbert space with the Dirac sense inner
product, it is Hermitian in a Hilbert space endowed by the inner product
$\langle n|\eta|m\rangle$ as well as having a correct-sign propagator (no
Lee-Wick fields). Besides, the obtained metric operator also diagonalizes the
Hamiltonian in the two fields ( no mixing). Moreover, the Hermiticity of
$\eta$ constrained the two Higgs masses to be related as $M>2m$, which has
been obtained in another work using a very different regime and thus supports
our calculation. Also, an equivalent Hermitian (in the Dirac sense)
Hamiltonian is obtained which has no Ghost states at all, which is a forward
step to make the Lee-Wick theories more popular among the Physicists.

\end{abstract}
\maketitle

One of the greatest puzzles in Particle Physics is the Hierarchy problem
\cite{DJOUADI}. SUSY was invented to solve this problem \cite{susy}. However,
very recently and on the guidance of a previous work of Lee and Wick
\cite{lee1,lee2}, a Lee-Wick (LW) extension of the standard model has been
introduced and investigated \cite{lee-wick,caron,LHD,neutrino}. While the LW
QED is a finite theory, the non-Abealian LW gauge theory is not finite.
Although it is not finite, it has been shown that it solves the Hierarchy
problem too.

The main idea of any Lee-Wick model is that the regulator in Pauli-Villars
corresponds to a physical degree of freedom. However, a \ QED with a Photon
propagator with the regulator term is a theory with higher derivative. A great
puzzle that makes such trends in the theory of particle physics not popular is
that they include exotic fields called the Lee-Wick fields. For instance, in
the LW extension of standard model every field of the conventional standard
model has a higher derivative kinetic term and it has been shown that the
theory can be converted into an equivalent one with more fields but some of
them has a propagator with wrong sign (exotic).

In a very different kind of studies, Carl Bender and Philip D Mannheim have
shown that a quantum mechanical theory with higher derivatives which
apparently suffers from negative norm problem can be converted into an
equivalent one with the ghost states are disappeared \cite{noghost}. In
showing that, they stressed a higher derivative Pais-Uhlenbeck model. In fact,
the regime of $\mathcal{PT}$ -symmetric theories has been used successfully in
some other works \cite{abophi41p1,abophi61p1,bendr}. In this letter, we show
that the ideas can successfully applied to the different sectors in the LW
Standard model introduced very recently. For that, we shall stress a type of
scalar field theory very similar to that employed\ in the Lee-Wick standard
model. We show that the theory is free from ghost states which then leads to
the enhancement of the popularity of the Lee-Wick standard model as a possible
theory free from ghosts as well as does not suffer from the Hierarchy puzzle.

A prototype scalar field Lagrangian introduced in the Lee-Wick standard model
is \cite{lee-wick};
\begin{equation}
\mathcal{L}=\frac{1}{2}\partial_{\mu}\phi\partial^{\mu}\phi-\frac{1}{2M^{2}%
}\left(  \partial^{2}\phi\right)  ^{2}-\frac{1}{2}m^{2}\phi^{2}.
\end{equation}
Fellowing the work in Ref. \cite{lee-wick}, one can introduce an auxiliary
field $\phi_{2}$ to get rid of the higher derivative in the theory such that;%

\begin{equation}
\mathcal{L}=\frac{1}{2}\partial_{\mu}\phi\partial^{\mu}\phi-\frac{1}{2}%
m^{2}\phi^{2}-\phi_{2}\partial^{2}\phi+\frac{1}{2}M^{2}\phi_{2}^{2},
\label{lag}%
\end{equation}
From the equation of motion of $\phi_{2}$ we get;%
\[
\frac{\partial\mathcal{L}}{\partial\left(  \partial_{\mu}\phi_{2}\right)
}=0,\text{ \ \ \ \ \ }\frac{\partial\mathcal{L}}{\partial\phi_{2}}=\phi
_{2}M^{2}-\partial^{2}\phi.\text{\ }%
\]
Then, the auxiliary field $\phi_{2}$ is given by the relation
\[
\phi_{2}=\frac{1}{M^{2}}\partial^{2}\phi.
\]
Let us define
\[
\phi=\phi_{1}-\phi_{2},
\]

Then;
\begin{align}
\mathcal{L}  &  =\frac{1}{2}\partial_{\mu}\left(  \phi_{1}-\phi_{2}\right)
\partial^{\mu}\left(  \phi_{1}-\phi_{2}\right)  -\frac{1}{2}m^{2}\left(
\phi_{1}-\phi_{2}\right)  ^{2}-\phi_{2}\partial^{2}\left(  \phi_{1}-\phi
_{2}\right)  +\frac{1}{2}M^{2}\phi_{2}^{2},\nonumber\\
&  =\frac{1}{2}\partial_{\mu}\phi_{1}\partial^{\mu}\phi_{1}+\frac{1}%
{2}\partial_{\mu}\phi_{2}\partial^{\mu}\phi_{2}-\partial_{\mu}\phi_{2}%
\partial^{\mu}\phi_{1}-\phi_{2}\partial^{2}\phi_{1}+\phi_{2}\partial^{2}%
\phi_{2}\label{lagnonh}\\
&  -\frac{1}{2}m^{2}\left(  \phi_{1}-\phi_{2}\right)  ^{2}+\frac{1}{2}%
M^{2}\phi_{2}^{2},\nonumber\\
&  =\frac{1}{2}\partial_{\mu}\phi_{1}\partial^{\mu}\phi_{1}-\frac{1}%
{2}\partial_{\mu}\phi_{2}\partial^{\mu}\phi_{2}-\frac{1}{2}m^{2}\phi_{1}%
^{2}+\frac{1}{2}\left(  M^{2}-m^{2}\right)  \phi_{2}^{2}\allowbreak+m^{2}%
\phi_{2}\phi_{1}.\nonumber
\end{align}
The Hamiltonian corresponding to the Lagrangian in Eq.(\ref{lagnonh}) can be
obtained as;
\begin{equation}
H=\frac{\pi_{1}^{2}}{2}+\frac{1}{2}\left(  \nabla\phi_{1}\right)  ^{2}%
+\frac{1}{2}m^{2}\phi_{1}^{2}-\frac{\pi_{2}^{2}}{2}-\frac{1}{2}\left(
\nabla\phi_{2}\right)  ^{2}-\frac{1}{2}\left(  M^{2}-m^{2}\right)  \phi
_{2}^{2}-m^{2}\phi_{1}\phi_{2} \label{hamh}%
\end{equation}

Now, let us apply the canonical transformation $\phi_{2}\rightarrow i\phi
_{2},\pi_{2}\rightarrow-i\pi_{2}$ which preserve the commutation relation
\cite{noghost};%
\begin{equation}
\left[  \phi_{2}\left(  x\right)  ,\pi_{2}\left(  y\right)  \right]  =\left[
i\phi_{2}\left(  x\right)  ,-i\pi_{2}\left(  y\right)  \right]  =i\delta
^{3}\left(  x-y\right)  .
\end{equation}

Then the transformed Hamiltonian will take the form;

\bigskip%
\begin{equation}
H=\frac{\pi_{1}^{2}}{2}+\frac{1}{2}\left(  \nabla\phi_{1}\right)  ^{2}%
+\frac{1}{2}m^{2}\phi_{1}^{2}+\frac{\pi_{2}^{2}}{2}+\frac{1}{2}\left(
\nabla\phi_{2}\right)  ^{2}+\frac{1}{2}\left(  M^{2}-m^{2}\right)  \phi
_{2}^{2}-im^{2}\phi_{1}\phi_{2} \label{hamnonh}%
\end{equation}
In other words, the negative norm manifested in the work of
Ref.\cite{lee-wick} by a negative kinetic term of the LW field ($\phi_{2}$) is
manifested here by the non-Hermiticity of the theory represented by the
Lagrangian in Eq. (\ref{lagnonh}). By assuming that $\phi_{2}$ is a pseudo
scalar, the Hamiltonian obtained from Eq. (\ref{lagnonh}) is $\mathcal{PT}%
$-symmetric too. Indeed, non-Hermitian $\mathcal{PT}$-symmetric theories are
suffering from the existence of ghost states however there exists known
algorithms to recover such problems\cite{spect,spect,noghost}. In fact, though
the Hamiltonian in Eq.(\ref{hamnonh}) is non-Hermitian in the Dirac sense, it
is not only Hermitian in a Hilbert space endowed by the inner product $\langle
n|\eta|m\rangle$ \cite{spect,spect1}, where $\eta$ is a positive definite
metric operator, but also the kinetic terms have the correct form.

To start the algorithm of curing the ghost states problem in the theory, for
simplicity, let us investigate, first, the theory in $0+1$ dimensions (Quantum
mechanics). Since the Hamiltonian in Eq.(\ref{hamnonh}) is pseudo-Hermitian,
one can seek a positive definite metric operator of the form;
\[
\eta=\exp\left(  2\left(  \omega_{1}\pi_{1}\phi_{2}+\omega_{2}\pi_{2}\phi
_{1}\right)  \right)  ,
\]
where $\omega_{1}$ and $\omega_{2}$ are two real parameters to be obtained
later in terms of the mass parameters $m$ and $M$. Note that $\eta$ is
Hermitian and has the property \cite{spect,spect1}
\begin{equation}
\eta H\eta^{-1}=H^{\dagger}.
\end{equation}
Also, $\rho=\sqrt{\eta}$ has the property
\begin{equation}
\rho H\rho^{-1}=h,
\end{equation}
where $h$ is a Hermitian (in the Dirac sense) as well as positive normed
Hamiltonian equivalent to $H$. \ 

To determine the parameters $\omega_{1}$ and $\omega_{2}$, we consider the
transformations of the different fields in the Hamiltonian under the effect of
$\rho$ as follows;
\begin{align*}
\rho\phi_{1}\rho^{-1}  &  =\phi_{1}-i\omega_{1}\phi_{2},\\
\rho\pi_{1}\rho^{-1}  &  =\pi_{1}+i\omega_{2}\pi_{2},\\
\rho\phi_{2}\rho^{-1}  &  =\phi_{2}-i\omega_{2}\phi_{1},\\
\rho\pi_{2}\rho^{-1}  &  =\pi_{2}+i\omega_{1}\pi_{1}.
\end{align*}
Accordingly;%
\begin{align*}
h  &  =\frac{\left(  \pi_{1}+i\omega_{2}\pi_{2}\right)  ^{2}}{2}+\frac{1}%
{2}m^{2}\left(  \phi_{1}-i\omega_{1}\phi_{2}\right)  ^{2}+\frac{\left(
\pi_{2}+i\omega_{1}\pi_{1}\right)  ^{2}}{2}\\
&  +\frac{1}{2}\left(  M^{2}-m^{2}\right)  \left(  \phi_{2}-i\omega_{2}%
\phi_{1}\right)  ^{2}-im^{2}\left(  \phi_{1}-i\omega_{1}\phi_{2}\right)
\left(  \phi_{2}-i\omega_{2}\phi_{1}\right)  ,
\end{align*}
or%

\begin{align}
h  &  =\frac{1}{2}\pi_{1}^{2}+i\pi_{1}\omega_{2}\pi_{2}-\frac{1}{2}\omega
_{2}^{2}\pi_{2}^{2}+\frac{1}{2}m^{2}\phi_{1}^{2}-\frac{1}{2}m^{2}\omega
_{1}^{2}\phi_{2}^{2}+\frac{1}{2}\pi_{2}^{2}+i\pi_{2}\omega_{1}\pi
_{1}\nonumber\\
&  -\frac{1}{2}\omega_{1}^{2}\pi_{1}^{2}\frac{1}{2}\left(  M^{2}-m^{2}\right)
\left(  \phi_{2}^{2}-\omega_{2}^{2}\phi_{1}^{2}\right)  -m^{2}\omega_{2}%
\phi_{1}^{2}-m^{2}\omega_{1}\phi_{2}^{2}-im^{2}\phi_{1}\phi_{2}\\
&  +im^{2}\omega_{1}\allowbreak\phi_{2}\omega_{2}\phi_{1}-i\phi_{2}\omega
_{2}\phi_{1}M^{2}+i\phi_{2}\omega_{2}\phi_{1}m^{2}-im^{2}\phi_{1}\omega
_{1}\phi_{2}.\nonumber
\end{align}
For $h$ to be Hermitian, one has to put the constraints
\begin{align}
i\omega_{2}+i\omega_{1}  &  =0,\nonumber\\
\left(  -m^{2}+m^{2}\omega_{1}\omega_{2}-\omega_{2}M^{2}+\omega_{2}m^{2}%
-m^{2}\omega_{1}\right)   &  =0,
\end{align}
on the introduced parameters $\omega_{1}$ and $\omega_{2}$. Equivalently, we
have the relations
\begin{align}
\omega_{1}  &  =-\omega_{2},\nonumber\\
-m^{2}-m^{2}\omega_{1}^{2}+\omega_{1}M^{2}-2m^{2}\omega_{1}  &  =0.
\end{align}
In terms of the mass parameters, $\omega_{1}$ can be obtained as%

\begin{equation}
\omega_{1}=\frac{1}{2m^{2}}\left(  M^{2}-2m^{2}\pm\sqrt{M^{4}-4M^{2}m^{2}%
}\right)  .
\end{equation}
Also, due to the reality of $\omega_{1}$, the two Higgs masses are related by;%
\[
M^{2}\geqslant4m^{2},
\]
which agrees with the results in Ref.\cite{lee-wick}

\begin{figure}[ptb]
\begin{center}
\includegraphics{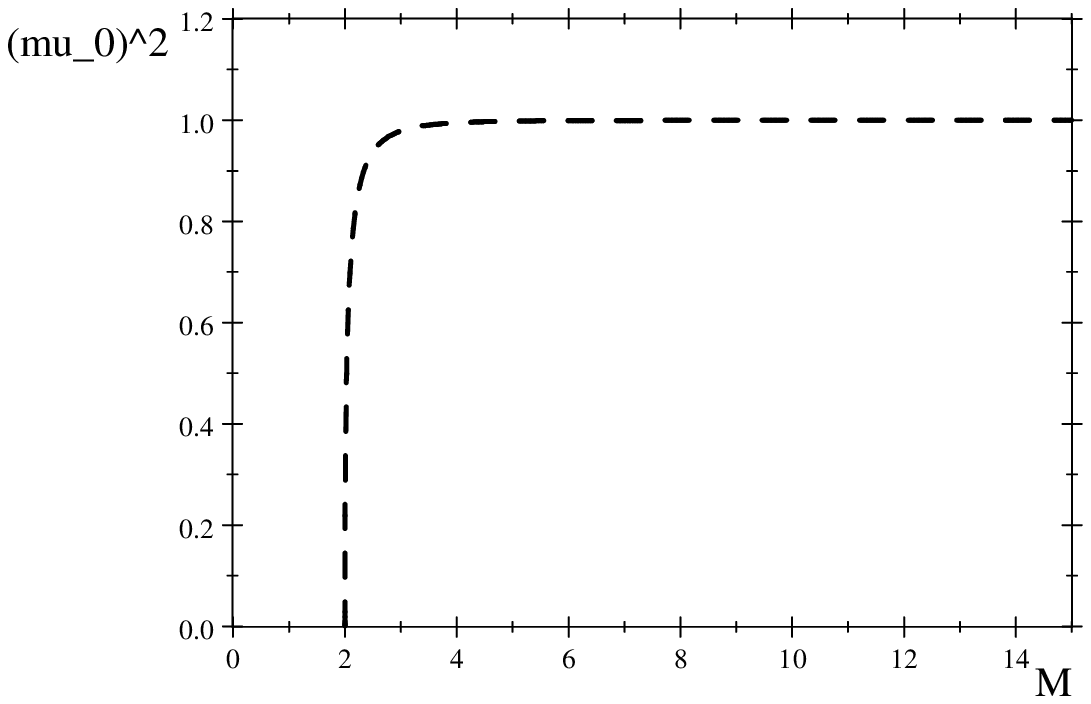}
\end{center}
\caption{The factor $\mu_{0}^{2}=(1-\omega^{2})$ plotted against the mass
parameter $M$ for $m=1$. One can realize that the factor is positive for the
available range of $M$.}%
\label{Leewick__1}%
\end{figure}

\begin{figure}[ptb]
\begin{center}
\includegraphics{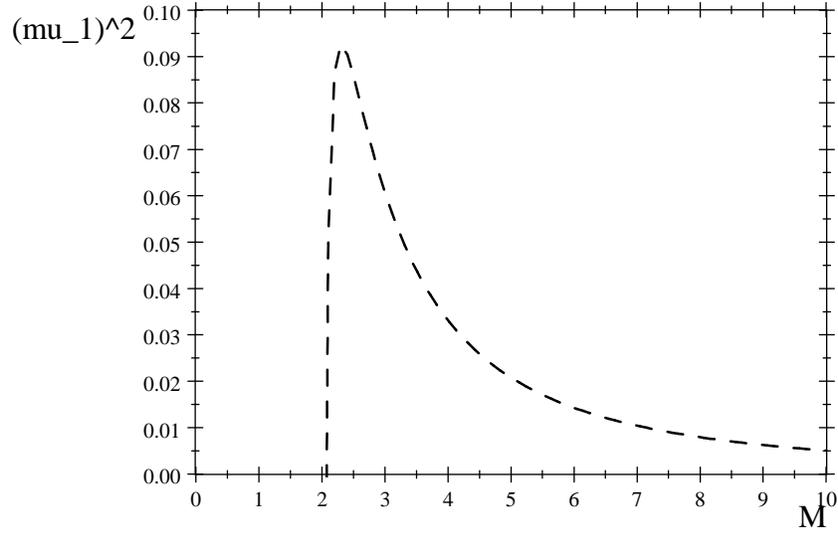}
\end{center}
\caption{A contribution to the mass parameter squared of the field $\phi_{1}$
of the form $\mu_{1}^{2}=\left(  \frac{1}{2}m^{2}\omega_{1}^{2}+m^{2}%
\omega_{1}-\frac{1}{2}M^{2}\omega_{1}^{2}\right)  $ in the Hermitian
Hamiltonian $h$, plotted against the mass parameter $M$ for $m=1$. Since the
other contribution is $m^{2}$ and from the plot $\mu_{1}^{2}$ is always
positive the mass squared as a whole is positive.}%
\label{Leewick__2}%
\end{figure}

\begin{figure}[ptb]
\begin{center}
\includegraphics{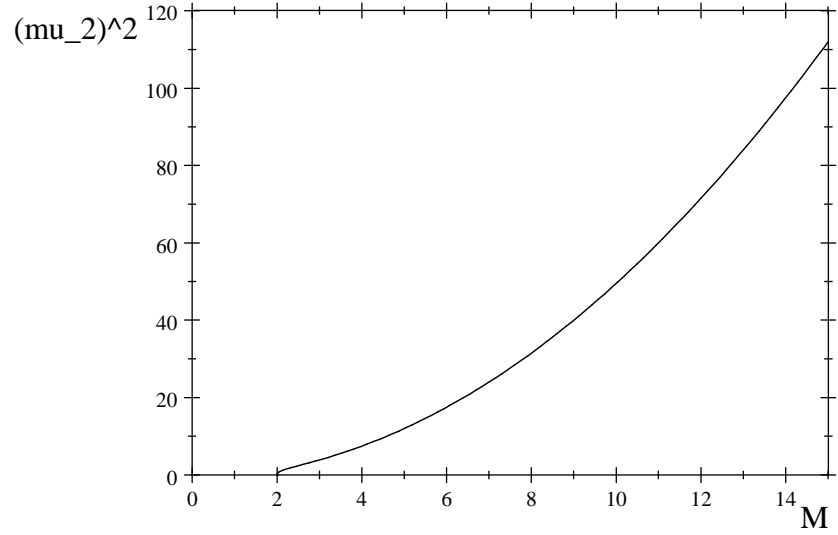}
\end{center}
\caption{The mass parameter squared of the field $\phi_{2}$ given by $\mu
_{2}^{2}=-\frac{1}{2}m^{2}\omega_{1}^{2}+\frac{1}{2}M^{2}-m^{2}\omega
_{1}-\frac{1}{2}m^{2}$ in the Hermitian Hamiltonian $h$, plotted against the
mass parameter $M$ for $m=1$, which is again a positive quantity.}%
\label{Leewick__3}%
\end{figure}

Then the Hermitian Hamiltonian $h$ has the form;
\begin{align*}
h  &  =\frac{1}{2}\pi_{1}^{2}\left(  1-\omega_{1}^{2}\right)  +\frac{1}%
{2}m^{2}\phi_{1}^{2}+\frac{1}{2}\left(  1-\omega_{1}^{2}\right)  \pi_{2}%
^{2}-\frac{1}{2}m^{2}\omega_{1}^{2}\phi_{2}^{2}\\
&  +\frac{1}{2}\left(  M^{2}-m^{2}\right)  \left(  \phi_{2}^{2}-\omega_{1}%
^{2}\phi_{1}^{2}\right)  +m^{2}\omega_{1}\phi_{1}^{2}-m^{2}\omega_{1}\phi
_{2}^{2},\\
&  =\frac{1}{2}\pi_{1}^{2}\left(  1-\omega_{1}^{2}\right)  +\frac{1}{2}%
m^{2}\phi_{1}^{2}+\frac{1}{2}\left(  1-\omega_{1}^{2}\right)  \pi_{2}^{2}\\
&  +\left(  \frac{1}{2}m^{2}\omega_{1}^{2}+m^{2}\omega_{1}-\frac{1}{2}%
M^{2}\omega_{1}^{2}\right)  \phi_{1}^{2}\\
&  +\left(  -\frac{1}{2}m^{2}\omega_{1}^{2}+\frac{1}{2}M^{2}-m^{2}\omega
_{1}-\frac{1}{2}m^{2}\right)  \phi_{2}^{2}.
\end{align*}
To make sure that the negative norm problem has been lifted, we plotted the
propagator-sign governing factors of the form $\mu_{0}^{2}=\left(
1-\omega_{1}^{2}\right)  $, $\ \mu_{1}^{2}=\left(  \frac{1}{2}m^{2}\omega
_{1}^{2}+m^{2}\omega_{1}-\frac{1}{2}M^{2}\omega_{1}^{2}\right)  $ and $\mu
_{2}^{2}=\left(  -\frac{1}{2}m^{2}\omega_{1}^{2}+\frac{1}{2}M^{2}-m^{2}%
\omega_{1}-\frac{1}{2}m^{2}\right)  $ as a function of $M$ for $m=1$, \ in
Fig.\ref{Leewick__1}., Fig.\ref{Leewick__2}. and Fig.\ref{Leewick__3},
respectively. In these plots, we have taken the root $\omega_{1}=\frac
{1}{2m^{2}}\left(  M^{2}-2m^{2}-\sqrt{M^{4}-4M^{2}m^{2}}\right)  $, while the
other root represents a theory of indefinite norm. One can realize that all
these factors are positive for the available range of $M$ which assures the
remedy of the wrong sign in the propagator of the LW field.

In higher dimensions (Quantum field theory), one needs to deal with operator
densities and thus the metric operator will take the from;
\[
\eta=\int d^{3}z\exp\left(  2\left(  \omega_{1}\pi_{1}\left(  z\right)
\phi_{2}\left(  z\right)  +\omega_{2}\pi_{2}\left(  z\right)  \phi_{1}\left(
z\right)  \right)  \right)  .
\]
Accordingly, we have the relations%
\begin{align*}
\rho\phi_{1}\left(  x\right)  \rho^{-1} &  =\phi_{1}\left(  x\right)
-i\omega_{1}\int d^{3}z\phi_{2}\left(  z\right)  \delta^{3}(x-z),\\
\rho\pi_{1}\rho^{-1} &  =\pi_{1}+i\omega_{2}\int d^{3}z\pi_{2}\left(
z\right)  \delta^{3}(x-z),\\
\rho\phi_{2}^{-1}\rho^{-1} &  =\phi_{2}-i\omega_{2}\int d^{3}z\phi_{1}\left(
z\right)  \delta^{3}(x-z),\\
\rho\pi_{2}^{-1}\rho^{-1} &  =\pi_{2}+i\omega_{1}\int d^{3}z\pi_{1}\left(
z\right)  \delta^{3}(x-z).
\end{align*}
And thus
\begin{align*}
\rho\phi_{1}^{-1}\rho^{-1} &  =\phi_{1}-i\omega_{1}\phi_{2},\\
\rho\pi_{1}^{-1}\rho^{-1} &  =\pi_{1}+i\omega_{2}\pi_{2},\\
\rho\phi_{2}^{-1}\rho^{-1} &  =\phi_{2}-i\omega_{2}\phi_{1},\\
\rho\pi_{2}^{-1}\rho^{-1} &  =\pi_{2}+i\omega_{1}\pi_{1}.
\end{align*}
Also, note that
\begin{align}
\rho\frac{1}{2}\left(  \nabla\phi_{1}(x)\right)  ^{2}\rho^{-1} &  =\frac{1}%
{2}\left(  \nabla\phi_{1}(x)\right)  ^{2}-i\omega_{1}\nabla_{x}\phi
_{1}(x)\nabla_{x}\phi_{2}(x)-\frac{\omega_{1}^{2}}{2}\left(  \nabla_{x}%
\phi_{2}(x)\right)  ^{2},\nonumber\\
\rho\frac{1}{2}\left(  \nabla\phi_{2}(x)\right)  ^{2}\rho^{-1} &  =\frac{1}%
{2}\left(  \nabla\phi_{2}(x)\right)  ^{2}-i\omega_{2}\nabla_{x}\phi
_{1}(x)\nabla_{x}\phi_{2}(x)-\frac{\omega_{2}^{2}}{2}\left(  \nabla_{x}%
\phi_{1}(x)\right)  ^{2}.
\end{align}
Again, with the choice $\omega_{1}=-\omega_{2}=\omega$, one gets;%

\begin{equation}
\rho\left(  \frac{1}{2}\left(  \nabla\phi_{1}(x)\right)  ^{2}+\frac{1}%
{2}\left(  \nabla\phi_{2}(x)\right)  ^{2}\right)  \rho^{-1}=\frac{1}{2}\left(
1-\omega^{2}\right)  \left(  \nabla\phi_{1}(x)\right)  ^{2}+\left(  \nabla
\phi_{2}(x)\right)  ^{2},
\end{equation}
and the quantum field Hermitian Hamiltonian takes the form;%
\begin{align}
h  &  =\frac{1}{2}\pi_{1}^{2}\left(  1-\omega^{2}\right)  +\frac{1}{2}\left(
1-\omega^{2}\right)  \left(  \nabla\phi_{1}(x)\right)  ^{2}+\frac{1}{2}%
m^{2}\phi_{1}^{2}\nonumber\\
&  +\frac{1}{2}\left(  1-\omega^{2}\right)  \pi_{2}^{2}+\frac{1}{2}\left(
1-\omega^{2}\right)  \left(  \nabla\phi_{2}(x)\right)  ^{2}+\frac{1}{2}%
M^{2}\phi_{2}^{2}\nonumber\\
&  +\left(  \frac{1}{2}m^{2}\omega^{2}+m^{2}\omega-\frac{1}{2}M^{2}\omega
_{1}^{2}\right)  \phi_{1}^{2}\label{Herm}\\
&  +\left(  -\frac{1}{2}m^{2}\omega_{1}^{2}-m^{2}\omega_{1}-\frac{1}{2}%
m^{2}\right)  \phi_{2}^{2}.\nonumber
\end{align}
One can easily realize that the governing factors are all positive for the
available range of the mass parameter $M$ relative to the mass parameter $m$.
Accordingly, the problem of wrong sign propagator has been recovered. Another
benefit of the transformation mapping $H\rightarrow h$, is that there exists
no mixing terms in $h$ ($h$ is diagonal in the fields $\phi_{1}$ and $\phi
_{2}$) .

To make sure that the Hermitian equivalent Hamiltonian in Eq.(\ref{Herm})
still bears the feature of quadratic divergence cancellation we rewrite it in
the form;%
\begin{align*}
h  &  =h_{1}+h_{2},\\
h_{1}  &  =\frac{1}{2}\left(  \pi_{1}^{2}+\left(  \nabla\phi_{1}\right)
^{2}\right)  +\frac{1}{2}m^{2}\left(  1+\omega_{1}\right)  ^{2}\phi_{1}^{2}\\
&  +\frac{1}{2}\left(  -\omega_{1}^{2}\right)  \left(  \pi_{2}^{2}+\left(
\nabla\phi_{2}\right)  ^{2}\right)  +\frac{1}{2}\left(  -m^{2}\left(
1+\omega_{1}\right)  ^{2}\right)  \phi_{2}^{2},\\
h_{2}  &  =\frac{1}{2}\left(  \pi_{2}^{2}+\left(  \nabla\phi_{2}\right)
^{2}\right)  +\frac{1}{2}M^{2}\phi_{2}^{2}+\frac{1}{2}\left(  -\omega_{1}%
^{2}\right)  \left(  \pi_{1}^{2}+\left(  \nabla\phi_{1}\right)  ^{2}\right) \\
&  +\frac{1}{2}\left(  -M^{2}\omega_{1}^{2}\right)  \phi_{1}^{2},
\end{align*}
which shows that although $h$ is Hermitian and positive normed it can be
decomposed into two terms each of which has the from of a normal and a
Lee-Wick fields.

\section*{Conclusions}

We considered a higher derivative scalar field theory of the form used in the
Lee-Wick standard model. We were able to obtain a non-Hermitian but
$\mathcal{PT}$-symmetric two-field equivalent Hamiltonian. Using the tools
applied to pseudo Hermitian Hamiltonians, we were able to obtain the positive
definite metric operator both on the quantum mechanical and quantum field
versions of the theory in a closed form. The so obtained equivalent Hermitian
Hamiltonian has propagators of correct sign which mean that the Ghost problem
has been cured. Moreover, the Hermitian Hamiltonian is diagonal in the fields.
Note that, we discarded the potential term as it is used to break the symmetry
and has no effect of the negative norm of the auxiliary field and thus one can
add it at the end to the equivalent Hermitian Hamiltonian we obtained. We
assert that the work presented here is fully new as it is the first time to
obtain the exact metric operator for a realistic quantum field theory. Also,
the idea here can be applied to all the sectors in the Lee-wick standard model
and thus obtain a theory which is non-SUSY, has no Ghosts, as well as solves
the Hierarchy problem. A note to be mentioned is that the mass of the
auxiliary field is greater than the normal Higgs which means that it is out of
any experimental tests carried out. We aim that in proving the non-existence
of Ghosts in a Lee-Wick theory we make those theories attract the attentions
of  researchers as they introduce finite Quantum Electrodynamics theory as
well as showing up a standard model free from the Hierarchy problem.

\begin{acknowledgments}
We would like to thank Dr. Shabaan Khalil for his help.
\end{acknowledgments}

\newpage

\end{document}